# Wideband microwave resonator for NV center vector magnetometry

S.V. Bolshedvorskii[1,2], E.A. Koreshin[3], V.V. Soshenko[1,2], I.S. Cojocaru[1,4], S.M. Drofa[1,2,4], P.G. Vilyuzhanina[1,4,5], N.I. Salangin[1,5], A.M. Kozodaev[1,4,5], A. Chernyavskiy[1,4,5], A.N. Smolyaninov[2], P.V. Kapitanova[3], S.Ya. Kilin[6,7] and A.V. Akimov[1,2,4*]

[1]*P.N. Lebedev Institute RAS, Leninsky Prospekt 53, Moscow, 119991, Russia*

[2]*LLC Diamond Sensors, Narvskaya st., 1A, Moscow, 125130, Russia*

[3]*Faculty of Physics and Engineering, ITMO University, Saint Petersburg, 197101, Russia*

[4]*Russian Quantum Center, Bolshoy Boulevard 30, building 1, Moscow, 143025, Russia*

[5]*Moscow Institute of Physics and Technology, 9 Institutskiy per., Dolgoprudny, Moscow Region, 141701, Russia*

[6]*National Research Nuclear University "MEPhI", 31, Kashirskoe Highway, Moscow, 115409, Russia*

[7]*B.I. Stepanov Institute of Physics NASB, 68, Nezavisimosty Ave, Minsk, 220072 Belarus*

email: a.akimov@rqc.ru

## ABSTRACT

The nitrogen-vacancy (NV) color center in diamond has found applications in magnetometry across various geometrical scales. In particular, vector magnetometry using hand-held devices with low power consumption is of interest for mobile and geological applications. However, such a device requires efficient driving of the four different orientations of the NV center, thus placing demands on a broadband microwave resonator that provides a uniform and highly insensitive microwave magnetic field at the resonance frequencies of the NV centers. Here, we propose a unique planar resonator design that enables efficient driving of all four orientations of NV color centers over a 100 MHz range with only 0.5 W of microwave input power and 250 mW of laser power, allowing vector magnetic field measurements. Using the planar resonator to drive NV centers in diamond, a compact vector magnetometer was fabricated and experimentally studied. The sensitivity of the device is about $1.2\ \mathrm{nT}/\sqrt{\mathrm{Hz}}$. While this is not a record sensitivity for diamond-based magnetometers, this device occupies its own niche as a compact hand-held magnetometer.

# I. INTRODUCTION

The nitrogen-vacancy (NV) color center in diamond has been the focus of significant interest over the last decades, since the detection of luminescence from single defects and the observation of optically detected magnetic resonance (ODMR) signals [1]. One of the active applications of this color center is quantum magnetometry [2]. Nowadays, commercially available NV quantum magnetometers include stand-alone magnetometers [3–8] and high-resolution NV magnetometers [9–13]. Efforts are also ongoing to achieve femtotesla-range sensitivity [14,15]. Various types of magnetometers have different sensitivity requirements. For biological applications, high sensitivity is often required, which may necessitate higher optical and microwave powers. In contrast, devices focused on spatial resolution prioritize resolution at the expense of sensitivity. Hand-held vector magnetometers are, to a large extent, limited by power consumption but may allow lower sensitivity if the vector nature of the field is measured [16,17].

The negatively charged NV center in diamond has a well-defined axis, with electron and nuclear spins of 1, for the $^{14}\mathrm{N}$ isotope considered here. Most NV magnetometers actively utilize the fact that NV fluorescence depends on the orientation of the electron spin parallel or perpendicular to the NV axis, $(m_s = \pm 1, m_s = 0)$, which leads to ODMR [18]. Vector NV-based magnetometry exploits the fact that a diamond crystal contains four distinct crystallographic orientations of the NV axes. Each orientation experiences a different projection of an external bias magnetic field, which lifts the degeneracy of the $m_s = \pm 1$ states via the Zeeman effect and produces a set of resonance frequencies that encode the vector components of the magnetic field [19]. Crucially, the Zeeman splitting also depends on temperature through the thermal shift of the zero-field splitting, separating states $m_s = 0$ and $m_s = \pm 1$. Therefore, tracking at least four resonance lines allows simultaneous extraction of the three vector components of the magnetic field and the temperature [14,19]. This sensing protocol requires a microwave magnetic field that can simultaneously address multiple transitions spread over a broad frequency range of 100 MHz. At the same time, the microwave magnetic field should be uniform across the diamond sensing volume to avoid systematic errors in the extracted field vector.

A variety of microwave delivery schemes have been employed to drive $m_s = 0 \rightarrow m_s = \pm 1$ spin transitions in NV centers. Narrowband designs, such as split-ring resonators [20] and dielectric resonators [21], can achieve very high-quality factors (up to $10^3 - 10^4$) and correspondingly low bandwidths. For example, a dielectric resonator with $Q \approx 2000$ at 2.87 GHz exhibits a -3 dB bandwidth of about 1.5 MHz. These resonators are power-efficient when driving a single resonance but are poorly suited for vector magnetic field measurements with NV centers. On the other hand, non-resonant microwave delivery schemes are widely used for broadband operation, including simple copper wires [22], coaxial loop coils [19,23], microstrip waveguides [24–27], and coplanar lines [28–30]. These schemes provide broad bandwidth and allow manipulation of NV centers at arbitrary frequencies, but they suffer from poor power efficiency and are not well suited for generating a uniform microwave magnetic field over a large volume.

An attractive middle ground is to design a planar resonator with a deliberately low Q-factor that provides a single broad resonance covering the required frequency interval. Such a design preserves a significant fraction of the power enhancement of a resonant structure while maintaining a bandwidth large enough to address all four NV orientations simultaneously. By engineering a low-Q broadband resonance, one can obtain a -3 dB bandwidth of 100 MHz centered near 2.87 GHz, which is sufficient for the majority of vector magnetometry protocols [8,19,31]. This regime allows efficient conversion of the input power into a magnetic field, with a tenfold or greater improvement in the Rabi frequency per $\sqrt{\mathrm{W}}$ compared with a non-resonant microwave delivery scheme. Furthermore, the field distribution can be optimized to achieve a large homogeneous region, a requirement often neglected in simple broadband loops, where the magnetic field decays sharply with distance. Thus, a properly engineered low-Q resonator can simultaneously provide broad bandwidth, high power efficiency, and spatial uniformity of the microwave magnetic field.

In this paper, we present the design, fabrication, and characterization of a broadband planar microwave resonator optimized for vector magnetometry with NV centers in diamond. The planar configuration was chosen for its inherent compatibility with standard printed circuit board technology, which ensures low manufacturing cost, high reproducibility, and straightforward integration into industrially produced devices. We systematically analyzed the trade-off between the Q-factor, bandwidth, magnetic field amplitude, and field uniformity over a selected volume. Three different planar resonator topologies were developed. Among these, one was shown to cover the frequency range from 2.72 to 2.95 GHz with a return loss below 1 dB and magnetic field nonuniformity of less than 5% across a cylindrical volume 2 mm in diameter and 0.4–0.6 mm in height. This planar resonator topology enabled simultaneous excitation of all four NV orientations under a 4.4 mT bias field. To demonstrate the resonator efficiency, a diamond-based dc NV magnetometer with a sensitivity of $1.2\ \mathrm{nT}/\sqrt{\mathrm{Hz}}$ was realized using a moderate microwave input power of 0.5 W and a laser power of 250 mW. The achieved sensitivity, while not record-breaking, represents a practical and well-justified choice for a compact hand-held device with relatively low power consumption.

## II. WIDEBAND MICROWAVE RESONATOR DESIGN

The main factors considered during the development of the planar resonator numerical models are as follows. The resonator is intended to generate the magnetic field in the region of the diamond plate, whose overall dimensions do not exceed 27 × 11 × 2 mm. The generated magnetic field must be oriented either horizontally or vertically and must have a strength of at least 40 A/m over the frequency range from 2.72 to 2.95 GHz. Numerical models for three resonator topologies were developed in CST Microwave Studio, and their characteristics were simulated to find the optimal topology. Rogers RO4350 laminate with a thickness of 0.5 mm was considered as the substrate material. During the numerical simulation, the losses in dielectric and metal materials were taken into account.

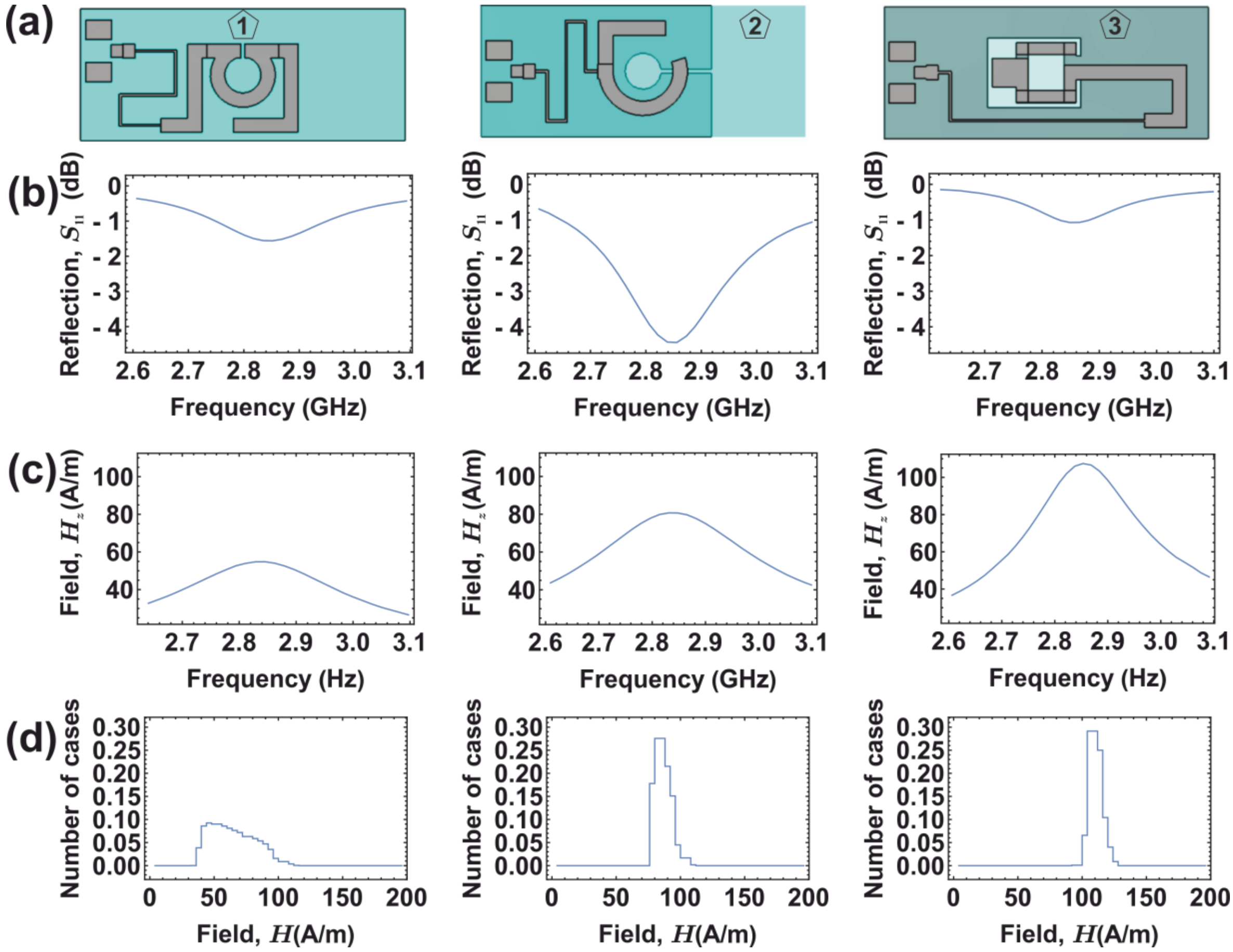

Figure 1. The resonator topologies and results of the numerical study of their characteristics. (a) Topology 1, 2, and 3 of the planar resonators. (b) Numerically simulated reflection coefficient of the three planar resonator topologies. (c) Numerically obtained magnetic field amplitude produced by topologies 1, 2, and 3 at 0.5 W of input power. (d) Histograms of magnetic fields in the cylindrical region with a diameter of 2 mm and a height of 0.4–0.6 mm over the surface of the three planar resonator topologies. The number of bins used is 50.

Topology 1 is based on a quarter-wave open-circuit stub connected in series with a ring (Figure 1a). The maximum current occurring at the end of such a stub is directed into the ring, creating a strong magnetic field at its center. With the appropriate selection of parameters (ring radius, strip width, and matching elements), this structure was numerically optimized within the required frequency band. Topology 2 is a conventional magnetic near-field loop, well known in the literature as an electrically small magnetic probe (Figure 1a). The symmetric design is not inherently resonant and requires matching to a 50 Ω transmission line using stubs and transformers. Topology 3 is a resonant circuit formed by a parallel-plate capacitor and a series inductor (Figure 1a). The key difference from the first two topologies is the greater number of degrees of freedom available for optimization in addition to the geometric dimensions. The capacitance $C$, inductance $L$, and, most importantly, the characteristic impedance ($L/C$ ratio) can be varied independently of the resonant frequency. This allows flexible control over the bandwidth and the amplitude of the magnetic field.

To excite the resonator topologies in CST Microwave Studio, a 50 Ω port was used. To match the input impedance of each resonator topology to 50 Ω, an impedance-matching network was designed. Figure 1b shows the numerically simulated reflection coefficients of planar resonator topologies 1, 2, and 3. Over the desired frequency band of 2.72–2.95 GHz, the best matching is obtained for topology 2, for which the reflection coefficient is better than −4 dB. Topologies 1 and 3 provide approximately −1 dB matching. During the numerical study, we also modeled the amplitude of the normal component of the magnetic field ($H_z$) at the center of the planar resonator at a distance of 0.5 mm from the substrate surface for all the topologies. The obtained results are shown in Figure 1c. The highest magnetic field amplitude, exceeding 100 A/m, is obtained for planar resonator topology 3. For topologies 1 and 2, the normal components of the magnetic field are 50 A/m and 80 A/m, respectively. To assess the uniformity of the magnetic field in the working volume above the planar resonator, the magnetic field distribution was numerically evaluated within a cylinder 2 mm in diameter and extending from 0.4 to 0.6 mm above the center of each planar resonator. The chosen volume makes it possible to work with large ensembles of NV centers during magnetic field measurements. The magnetic field distribution within the cylinder for each resonator topology is presented as a histogram in Figure 1d. The horizontal axis corresponds to the magnitude of the magnetic field normal component ($H_z$), expressed in amperes per meter (A/m), while the vertical axis represents the relative amplitude or frequency of occurrence $I(H_z)$, expressed in arbitrary units. Each histogram therefore represents the probability density function of the field magnitude within the cylindrical volume above the planar resonator surface.

To calculate the uniformity of the field, the weighted variance of the magnetic field distribution was calculated with the following equation:

$$\sigma_{H_z}^2 = \frac{\int I(H_z) \times (H_z - \langle H_z \rangle)^2}{\int I(H_z)}, \tag{1}$$

where $\langle H_z \rangle$ – the mean value of the simulated magnetic field $H_z$ in the cylinder over the chosen volume. Then the relative magnetic field uniformity $\epsilon$ is defined as a ratio of the weighted standard deviation to the weighted mean field:

$$\epsilon = \frac{\sigma_{H_z}}{\langle H_z \rangle} \times 100\% . \tag{2}$$

The calculated mean value of the magnetic field, weighted standard deviation, and relative uniformity for each planar resonator topology are summarized in Table 1.

Table 1. Weighted statistical parameters for the three resonator topologies.

| Planar resonator | $\langle H_z \rangle$ , A/m | $\sigma_{H_z}$ , A/m | $\epsilon$ , % |
|---|---|---|---|
| Topology 1 | 63.2 | 16.4 | 25.94 |
| Topology 2 | 85.2 | 6.5 | 7.62 |
| Topology 3 | 106.9 | 5.3 | 4.95 |

The results demonstrate that topology 3 of the planar resonator exhibits the highest value of $\langle H_z \rangle$ with the uniformity of 4.95%, compared to 7.62% for topology 2 and 25.94% for topology 1. Comparison of the simulated results shows that topology 3 provides the best combination of parameters: the highest magnetic field magnitude together with the lowest nonuniformity of the magnetic field distribution over the selected volume. Thus, topology 3 was selected as the most promising design for fabrication and further experimental investigation.

# III. RESULTS AND DISCUSSION

## Production and characterization of a wideband planar resonator

To confirm the results of the numerical study, planar resonator topology 3 was fabricated on a 0.5-mm-thick Rogers RO4350 laminate substrate ((Figure 2a). Eight samples of the topology 3 planar resonator were fabricated, differing in their impedance-matching circuits. This made it possible, first, to refine the numerical model by taking into account the actual substrate parameters (dielectric constant and loss tangent) and technological deviations and, second, to compensate for uncertainties associated with the planar resonator environment in the experimental setup.

The measurements were carried out using an Arinst VNA-PR1 vector network analyzer. At the first stage, the reflection coefficient was measured for all samples over the frequency range of 2.2–3.2 GHz. The dimensionless magnetic field amplitude generated by the planar resonator samples was extracted from the measured reflection coefficient using the following equation:

$$H = \sqrt{\left(1 - |S_{11}|^2\right)} \ , \tag{3}$$

where $|S_{11}|$ is the measured reflection coefficient magnitude.

(Figure 2b provides the extracted magnetic field magnitude with the indication of the bandwidth at the -3 dB level for each sample under test. As one can see, sample No. 6 demonstrates the magnetic field amplitude higher than 0.6 arb. un. in the frequency band 2.764–2.878 GHz covering the central frequency of 2.83 GHz (see Figure 1c). This sample was used in subsequent magnetometry experiments.

It is known that the Rabi frequency for an NV center is directly proportional to the amplitude of the microwave field in the resonator $\Omega_R \sim B_{MW}$ . Under the assumption that the broadening caused by the Rabi frequency is negligible compared to the natural linewidth of the ODMR resonances, the resonance contrast can be taken as proportional to the square of the Rabi frequency $\mathrm{C}_i \sim \Omega_R^2$ [32]. Therefore, to qualitatively evaluate the experimental efficiency of the microwave resonator for NV-center state manipulation, the square roots of the ODMR resonance contrasts $\sqrt{\mathrm{C}_i}$ were normalized and overlaid on the plot of the magnetic-field amplitude in the resonator (see (Figure 2c).

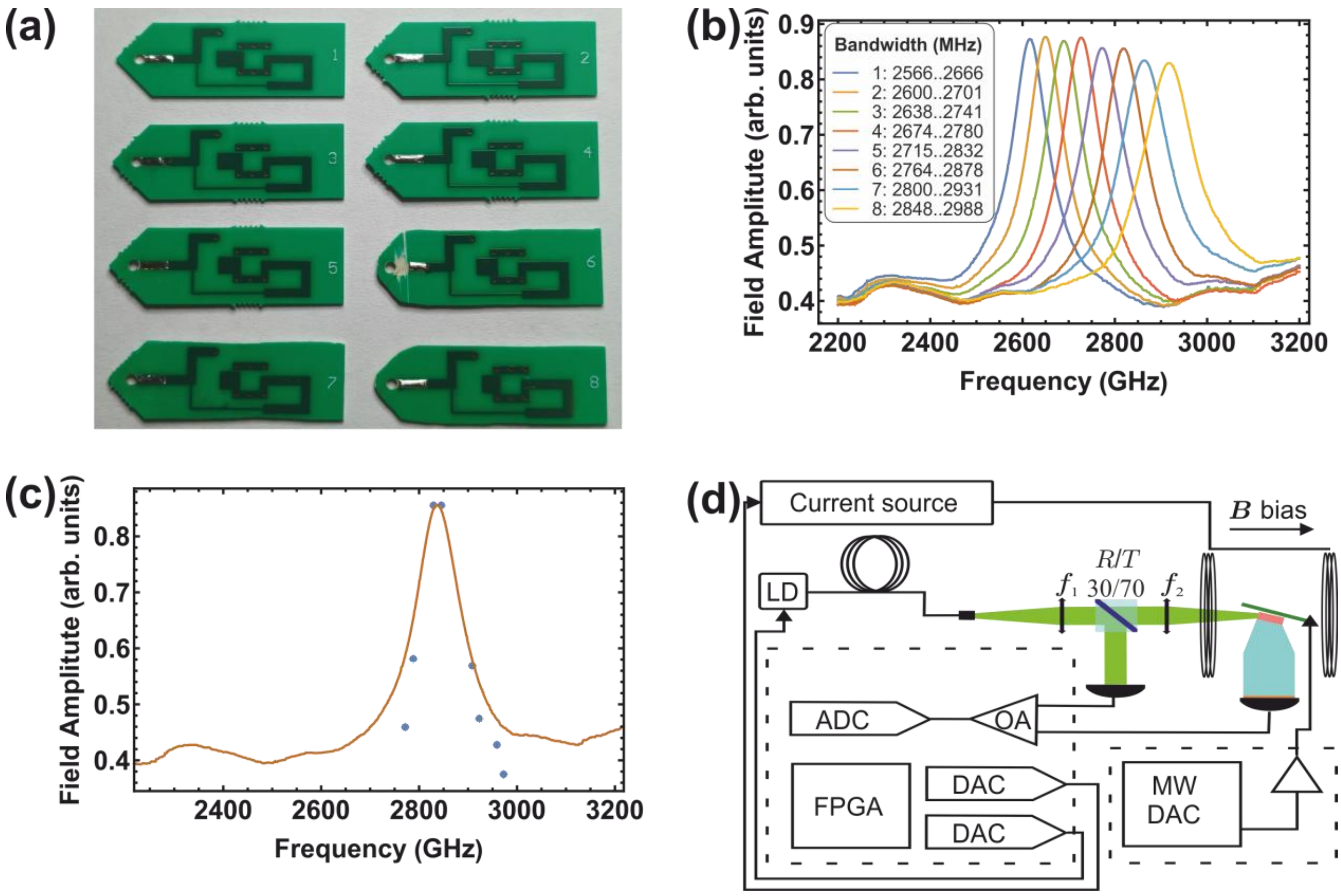

(Figure 2. a) Photograph of the fabricated samples of the planar resonator (topology 3). b) Relative magnetic field amplitude of the eight fabricated samples extracted from the measured reflection coefficient. The inset depicts the FWHM bandwidth for each sample. (c) Frequency profile of the microwave resonator obtained from the measured reflection coefficient (brown curve, $\sqrt{1-|S_{11}|^2}$ compared with the ODMR resonance amplitudes (blue points, $\sqrt{C_i}$ ). d) Experimental setup for ODMR measurements and magnetometry. OA – operational amplifier, ADC – analog-to-digital converter, DAC – digital-to-analog converter, FPGA – field-programmable gate array, $R/T$ – reflection-to-transmission ratio, LD – laser diode.

## Experimental setup

(Figure 2d depicts the experimental setup, including a diamond plate with NV centers, laser excitation, fluorescence detection, a bias magnetic field source, and microwave generation. The single-crystal diamond is a 1.2×1.2×0.35 mm$^3$ plate with a {100} front facet containing [$^{14}$N] ≈ 8 ppm, [NV$^-$] ≈ 0.9 ppm, and natural abundance of $^{13}$C atoms. The diamond is attached to a sapphire parabolic concentrator equipped with a 633 nm long-pass optical filter, which collects $NV^-$ fluorescence, suppresses laser light and unwanted scattering, and also provides thermal and mechanical stability. A 520 nm Wave Spectrum WSLX-520-200-M-H14-T-PD laser diode with a 105 µm optical-fiber output is used as the excitation source for the NV-center ensemble. The laser beam is collimated by an aspheric lens F1 (Thorlabs C105TMD-A). Then, 30% of the laser radiation is deflected by a beamsplitter (Thorlabs BS046) toward the reference photodiode, while the aspheric lens F2 (Thorlabs A397TM-A) focuses the laser light onto the diamond plate. The NV-ensemble signal, after optical noise cancellation similar to that described in [33], is digitized by an analog-to-digital converter (ADC). Helmholtz coils driven by a current source are used to apply a bias magnetic field $B_{bias}$, which is required to distinguish the four NV-center orientations associated with different crystallographic axes. The microwave signal is generated by a digital-to-analog converter and then amplified before being delivered to sample No. 6 of the planar resonator. The setup is controlled by a field-programmable gate array (FPGA) with digital-to-analog converters (DACs), which allow the laser-diode current and Helmholtz-coil current to be set and enable measurements of ODMR spectra and the magnetic field.

## ODMR spectra measurements

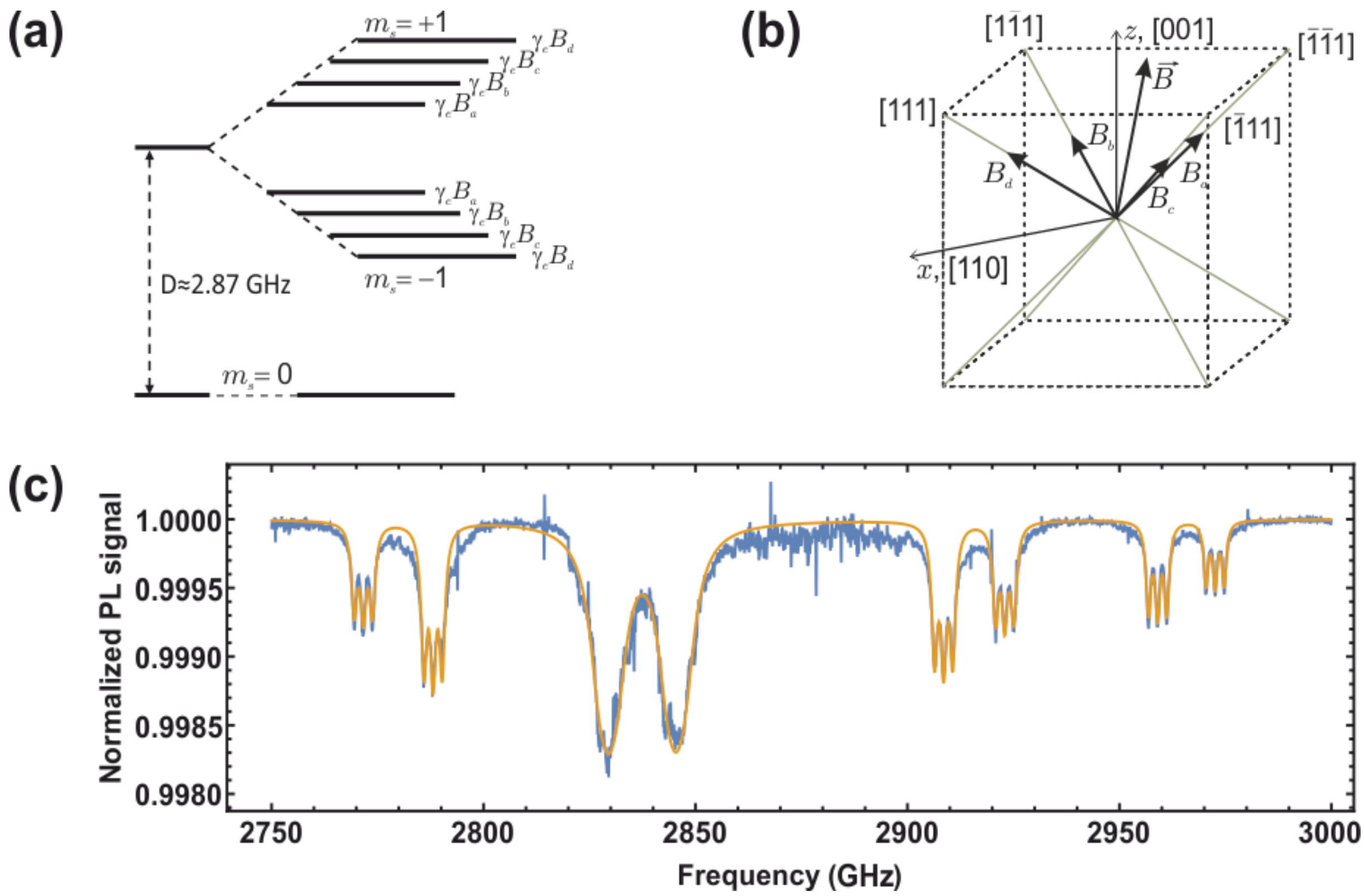

(Figure 3. a) Energy level diagram for the NV center ground state, with the zero-field splitting (ZFS) parameter $D$ and Zeeman shifts for $m_s = \pm 1$ energy levels in the presence of a bias magnetic field $\vec{B}$ for different projections of $\vec{B}$ along the NV symmetry axes. b) A laboratory-frame coordinate system $x, y, z$ defined in terms of diamond lattice vectors. A magnetic field $\vec{B}$ projects onto the four NV orientations $B_a$, $B_b$, $B_c$, $B_d$ causing the Zeeman shift shown in a. c) ODMR spectra show the PL signal from an ensemble of NV centers in a bias magnetic field $\vec{B}_0 = (1.51, 0.84, 4.08)$ mT, $|\vec{B}_0| = 4.43$ mT .

The negatively charged nitrogen-vacancy (NV⁻) center in diamond is a spin-1 system whose ground-state electronic spin sublevels ( $m_s = 0$ and $m_s = \pm 1$ ) are separated by a zero-field splitting of $D \approx 2.87$ GHz . In the presence of an external magnetic field $\vec{B}$ , the $m_s = \pm 1$ levels experience a Zeeman shift (see (Figure 3a), lifting their degeneracy and resulting in eight distinct ODMR resonances for an ensemble with four crystallographic orientations [34]. Each resonance corresponds to a spin transition $m_s = 0 \rightarrow m_s = \pm 1$ for a specific NV axis, with hyperfine splitting due to the coupling to the $^{14}\mathrm{N}$ nuclear spin ($I = 1$) [35] .

Using sample No. 6 of the planar resonator, we experimentally characterized the spin-manipulation efficiency by measuring the ODMR spectra of an ensemble of NV centers in diamond at a fixed microwave power of 19 dBm. First, the diamond plate containing the NV-center ensemble was placed in the homogeneous field region of the resonator. The planar resonator was aligned parallel to the front facet of the diamond crystal to provide an equivalent projection of the microwave magnetic field generated by the resonator onto each of the NV-center axes in the diamond plate. Second, a static bias magnetic field was applied to resolve all eight electron-spin transitions through the Zeeman effect (see (Figure 3c).

To extract the resonance frequencies and amplitudes, we performed a multi-peak fit of the measured ODMR spectrum. Each of the eight electron spin transitions was modeled as a triplet of Lorentzian peaks, accounting for the hyperfine interaction with the $^{14}\mathrm{N}$ nucleus [8]. The hyperfine splitting between the three components was fixed at $\Delta\nu_{hf} = 2.15\ \mathrm{MHz}$ , consistent with the known value for NV centers in diamond [36]. The experimental data were approximated with the following model:

$$S(f) = 1 + \sum_i \sum_j C_i \times \left( \Gamma_i^2 / \left( \left( f - f_{ij} \right)^2 + \Gamma_i^2 \right) \right), \quad (4)$$

where $i = 1,...,8$ are the indexes of the eight transitions, $j = 1,2,3$ are the indexes of the hyperfine components, $C_i$ are the amplitudes, $\Gamma_i$ are the linewidths, and $f_{ij} = f_i + (j-2) \times \Delta\nu_{hf}$ are the peak center frequencies. The baseline was fixed at 1, corresponding to the normalized off-resonance PL signal. The parameters $\{f_i, C_i, \Gamma_i\}$ were optimized using a least-squares fitting procedure, yielding the eight resonance line centers $f_i$ and eight amplitudes $C_i$.

The obtained resonance frequencies $f_i$ were then used to determine the external magnetic field vector $\vec{B}_0$ and the zero-field splitting $D$. The calculations were performed in a coordinate system defined with respect to the diamond crystal, as depicted in (Figure 3b. The unit vectors $\vec{e}_x$, $\vec{e}_y$, and $\vec{e}_z$ are chosen along the diamond crystallographic directions $[110]$, $[1\bar{1}0]$, and $[001]$, respectively. In this frame, the four NV symmetry axes are given by:

$$\begin{aligned}\vec{n}_d &= \left(\sqrt{\frac{2}{3}}, 0, \sqrt{\frac{1}{3}}\right) \parallel [111] \\ \vec{n}_a &= \left(0, -\sqrt{\frac{2}{3}}, -\sqrt{\frac{1}{3}}\right) \parallel [1\bar{1}\bar{1}] \\ \vec{n}_c &= \left(-\sqrt{\frac{2}{3}}, 0, \sqrt{\frac{1}{3}}\right) \parallel [\bar{1}\bar{1}1] \\ \vec{n}_b &= \left(0, \sqrt{\frac{2}{3}}, -\sqrt{\frac{1}{3}}\right) \parallel [11\bar{1}]\end{aligned} \tag{5}$$

The spin Hamiltonian of the NV⁻ ground state in the presence of a magnetic field $\vec{B}$ [37] can be written in the following way:

$$H_s = DS_z^2 + g_s \mu_B \vec{B} \times \vec{S} + QI_z^2 + A_\parallel S_z I_z + A_\perp \left(S_x I_x + S_y I_y\right), \tag{6}$$

where $S$ is the electronic spin and $I$ is the nuclear spin. The term $DS_z^2$ represents the zero-field splitting, $g_s \mu_B \vec{B} \times \vec{S}$ the electron Zeeman interaction, and $QI_z^2$ the nuclear quadrupole shift. The hyperfine coupling is separated into longitudinal $A_\parallel S_z I_z$ and transverse $A_\perp \left(S_x I_x + S_y I_y\right)$ parts, with the $z$-axis taken along the NV symmetry axis. The nuclear spin Zeeman interaction is neglected for simplicity.

For a quantitative analysis, the Hamiltonians were numerically diagonalized using the Quantum Toolbox in Python (QuTiP). The eigenvalues of $H$ give the energies of the nine spin sublevels. The eight resonance frequencies were then computed as the energy differences between the eigenstates predominantly corresponding to the $m_s = 0$ and $m_s = \pm 1$ manifolds, averaged over the nuclear spin states.

A numerical least-squares approximation of the Hamiltonian parameters to the eight experimentally determined frequencies $f_i$ give the three Cartesian components of the magnetic field $\left(B_x, B_y, B_z\right)$ and the zero-field splitting $D$. This procedure yielded a determination of the bias magnetic field $\vec{B}_0 = (1.51, 0.84, 4.08)$ mT, $|\vec{B}_0| = 4.43$ mT.

## Experiments on magnetometry and sensitivity assessments

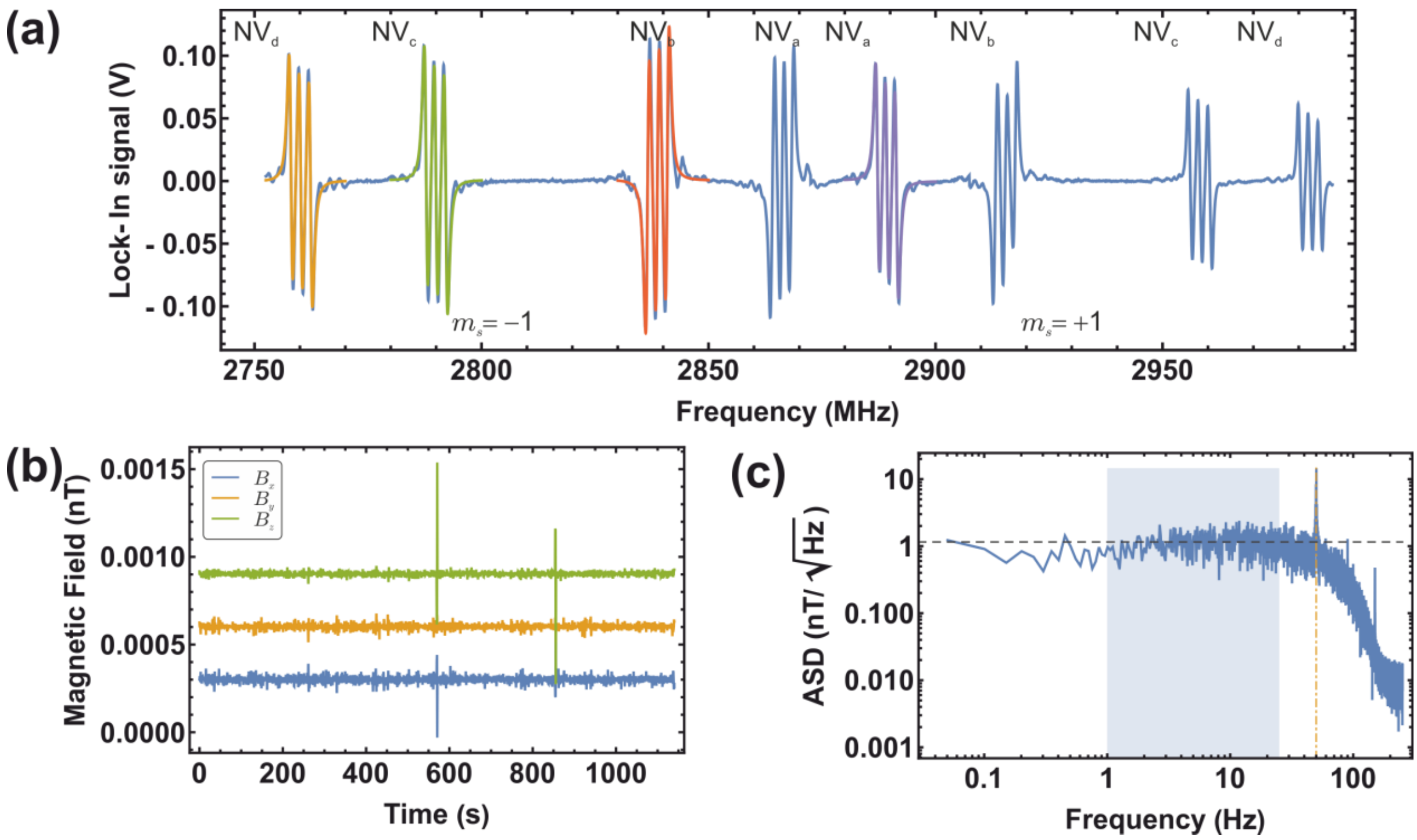

(Figure 4. a) Lock-in signal from the ensemble of NV centers. b) Time trace for the measured magnetic field. The average is subtracted for every plot, and the plots are artificially offset by 15 standard deviations. c) Amplitude spectral density (ASD) for the lock-in error signal. The vertical line stands for 50 Hz, and the horizontal line for the mean value over the first 25 Hz. The grey area stands for the range 1–25 Hz.

To demonstrate the practical performance of the developed wideband microwave resonator, vector magnetic field measurements were carried out using an ensemble of NV centers in diamond. A variety of methods are used in NV-center magnetometry experiments, including frequency modulation [38–40] and phase modulation [41–44] to address an electron-spin transition and encode information about the spin transitions within a frequency band around the modulation frequency. Here, we used the simultaneous broadband vector magnetometry protocol introduced by the Walsworth group [19], in which information from multiple NV orientations is encoded in separate frequency bands and measured using a single optical detector.

The four ODMR peaks corresponding to different NV crystallographic orientations were selected for simultaneous interrogation (see (Figure 4A). The selected transitions $m_s = 0 \rightarrow m_s = \pm 1$ for $NV_a$, $NV_b$, $NV_c$, $NV_d$ subensembles are colored in (Figure 4. These resonances were chosen because they enable simultaneous broadband vector magnetometry and lie within the operating frequency range of the microwave resonator under the applied bias magnetic field $\vec{B}_0$.

For each selected resonance, the microwave carrier frequency $f_c$ was frequency-modulated with a depth of 0.6 MHz at a distinct modulation frequency $f\,\mathrm{mod}\,i$. The modulation frequencies were chosen to achieve similar signal contrast while ensuring that none was an integer multiple of another, thereby avoiding crosstalk arising from nonlinear responses at harmonics.

The photoluminescence (PL) signal from all NV centers was collected on a single photodiode and digitized. Multi-channel lock-in demodulation was performed in software, extracting the in-phase component of the signal at each modulation frequency $f_i$. The demodulated signal $S_i$ for each channel is approximately proportional to the derivative of the ODMR line: $S_i \sim \frac{d}{d\mathrm{f}} PL(f)$.

To convert the demodulated signals $S_i$ into frequency shifts $\Delta f_i$, the lock-in signal slope $dS_i / df$ was measured for each channel. This was accomplished by sweeping the microwave carrier frequency around each resonance. The measured lock-in signal corresponds to the derivative of the ODMR lineshape. Consequently, the experimental data for each channel were fitted to a model consisting of a sum of three derivatives of Lorentzian functions with equal weights and equal hyperfine splitting $\Delta_{hf}$:

$$S_i(f) = \frac{1}{3} \sum_{j=-1}^{1} \left[ -2A_i \Gamma_i^2 \frac{f - \left(f_{0i} + j\Delta_{hf}\right)}{\left(\Gamma_i^2 + \left(f - \left(f_{0i} + j\Delta_{hf}\right)\right)^2\right)^2} \right], \quad (7)$$

where $A_i$ is the amplitude, $\Gamma_i$ is the linewidth, and $f_{0i}$ is the central frequency of the hyperfine triplet. The slope $dS_i / d$f was then evaluated as the derivative of this model at $f_{0i}$. The measured slopes and parameters detailing the vector magnetometer implementation are summarized in Table 2.

Table 2. Parameters detailing the implementation of simultaneous vector magnetometry.

| NV axis | Central carrier frequency $f_{0i}$, MHz | Unit vector | Modulation frequency $f\ \mathrm{mod}\, i$ , kHz | Slope $\lvert dS_i / df \rvert$, V/MHz |
|---|---|---|---|---|
| $\mathrm{NV}_a$ | 2889.3 | $\vec{n}_a \parallel \left[1\,\bar{1}\,\bar{1}\right]$ | 4.5 | 0.38 |
| $\mathrm{NV}_b$ | 2838.74 | $\vec{n}_b \parallel \left[1\,1\,\bar{1}\right]$ | 3.81 | 0.46 |
| $\mathrm{NV}_c$ | 2789.9 | $\vec{n}_c \parallel \left[1\,\bar{1}\,1\right]$ | 3.55 | 0.41 |
| $\mathrm{NV}_d$ | 2760.1 | $\vec{n}_d \parallel [111]$ | 2.32 | 0.35 |

The magnetic field in the laboratory frame $\vec{e}_x, \vec{e}_y, \vec{e}_z$ $\vec{B}(t) = \vec{B}_0 + \delta\vec{B}(t)$ is defined by measuring the frequency shifts $\Delta f_i(t)$ in each lock-in channel, which are related to the time-dependent field $\delta\vec{B}(t)$ through a linear transformation $L$. To establish this relationship, the matrix elements $L_{ij} = \frac{\partial f_i}{\partial B_j}$ are calculated as the derivatives of the resonance frequencies with respect to each magnetic field component at $\vec{B}_0$ and $D$ determined from ODMR measurements. The resulting matrix $L_{ij}$ is used to reconstruct the magnetic field vector in the laboratory frame using the following equation:

$$\vec{B}(t) = \vec{B}_0 + L_{ij}^{+} \Delta f_i(t), \tag{8}$$

where $L_{ij}^{+}$ is the pseudoinverse matrix. This approach is standard in NV-based vector magnetometry and is described in detail in the work [19] . The time trace for the magnetic field measured in a laboratory for over 1000 seconds is depicted in (Figure 4B.

The sensitivity of the NV magnetometer was evaluated from the amplitude spectral density (ASD) of the lock-in signal $S_i$. The time traces $S_i(t)$ were recorded for each channel, and ASD was computed using a fast Fourier transform (FFT). The mean magnetic field noise spectral density $\eta_B(f)$ was calculated as:

$$\eta_B(f) = \frac{ASD\, S_i(f)}{dS_i / dB}, \tag{9}$$

where $ASD\, S_i(f)$ is the amplitude spectral density of the lock-in signal, and $dS_i / dB$ is the sensitivity of the lock-in signal to the magnetic field. The latter is obtained from the slope $dS_i / df$ and the Zeeman shift coefficient $df / dB$ for the corresponding transition. The measured sensitivity was found to be $1.2\ \mathrm{nT}/\sqrt{\mathrm{Hz}}$ at frequencies above 1–25 Hz.

# IV. CONCLUSION

Motivated by the development of a compact hand-held magnetometer, three topologies of a broadband planar resonator were proposed to enable efficient microwave control of the internal states of NV centers. The resonator configured as an oscillatory circuit with a parallel-plate capacitor and a series inductor (topology 3) was selected as the best candidate, fabricated, and characterized. The microwave magnetic field amplitude of the planar resonator, measured at an input power of only 0.5 W, was 106.9 A/m at a central frequency of 2.83 GHz. Over a 114 MHz frequency band (2.764–2.878 GHz), the magnetic field amplitude nonuniformity was below 5%. Using the broadband planar resonator to drive NV centers in diamond, a compact vector magnetometer with a sensitivity of $1.2\ \mathrm{nT}/\sqrt{\mathrm{Hz}}$ was realized. The magnetometer operated in a continuous regime using synchronous detection of ODMR signals from NV centers in diamond. This technique enables continuous vector magnetic field measurements with a compact hand-held device.

# CONFLICT OF INTEREST

The authors have no financial/commercial conflicts of interest.

## ACKNOWLEDGMENT

The numerical study of the microwave planar resonator was supported by the Russian Science Foundation, project no. 25-42-10023 https://rscf.ru/project/25-42-10023/. The fabrication and characterization of the planar resonators and compact vector magnetometer were supported by a grant from the Ministry of Science and Higher Education of the Russian Federation No. 075-15-2024-556.

## DATA AVAILABILITY STATEMENT

The datasets used and/or analyzed during the current study are available from the corresponding author upon reasonable request.